\documentclass[useAMS]{mn2e}
\usepackage{amssymb,amsmath,psfig,times}
\usepackage{graphicx}
\def\gsim{ \lower .75ex \hbox{$\sim$} \llap{\raise .27ex \hbox{$>$}} }
\def\lsim{ \lower .75ex\hbox{$\sim$} \llap{\raise .27ex \hbox{$<$}} }

\def\beq{\begin{equation}}
\def\eeq{\end{equation}}

\def\sw{{\it Swift}}
\def\fe{{\it Fermi}}

\def\cgro{{\it CGRO}}

\def\ama{$E_{\rm p}-E_{\rm iso}$}

\def\th{$\theta_{\rm jet}$}

\def\thv{$\theta_{\rm view}$}

\def\G{$\Gamma_{0}$}

\title[Radio observations of GRBs]
{Radio afterglows of a complete sample of bright \sw\ GRBs: predictions from present days to the SKA era}
\author[G. Ghirlanda et al.]
{G. Ghirlanda$^{1}$\thanks{E--mail:giancarlo.ghirlanda@brera.inaf.it}, 
R. Salvaterra$^{2}$,   D. Burlon$^{3,4}$,  S. Campana$^{1}$,  A. Melandri$^{1}$, M. G. Bernardini$^{1}$, \newauthor 
S. Covino$^{1}$, P. D'Avanzo$^{1}$, V. D'Elia$^{5,6}$, G. Ghisellini$^{1}$, L. Nava$^{7,1}$, I. Prandoni$^{8}$, L. Sironi$^{9}$,  \newauthor 
G. Tagliaferri$^{1}$, S. D. Vergani$^{10,1}$, A. Wolter$^{1}$  \\
$^{1}$INAF -- Osservatorio Astronomico di Brera, via E. Bianchi 46, I-23807 Merate, Italy\\
$^{2}$INAF -- IASF Milano, via E. Bassini 15, I-20133 Milano, Italy \\
$^{3}$Sydney Institute for Astronomy, School of Physics, The University of Sydney, NSW 2006, Australia\\
$^{4}$ARC Centre of Excellence for All-sky Astrophysics (CAASTRO), The University of Sydney, NSW 2006, Australia\\
$^{5}$ASI -- Science Data Center, via Galileo Galilei, 00044 Frascati, Italy\\
$^{6}$INAF-OAR Via Frascati 33, I-00040 Monteporzio Catone Italy\\
$^{7}$APC Universit\'e Paris Diderot, 10 rue Alice Domon et Leonie Duquet, F-75205 Paris Cedex 13, France\\
$^{8}$INAF -- Istituto di Radioastronomia, via P. Gobetti, 101, I-40129 Bologna, Italy\\
$^{9}$ Harvard-Smithsonian Center for Astrophysics, 60 Garden Street, Cambridge, MA 02138, USA\\
$^{10}$ GEPI, Observatoire de Paris, CNRS, Univ. Paris Diderot, 5 place Jules Janssen, 92190, Meudon, France\\
}
\begin{document}

\date{}


\maketitle

\label{firstpage}

\begin{abstract}
Radio observations of Gamma Ray Bursts afterglows are fundamental in providing insights into their physics and environment, and in constraining the true energetics of these sources. Nonetheless, radio observations of GRB afterglows are presently sparse in the time/frequency domain. Starting from a complete sample of 58 bright \sw\ long bursts (BAT6), we constructed a homogeneous sub--sample of 38 radio detections/upper limits which preserves all the properties of the parent sample. One half of the bursts have detections between 1 and 5 days after the explosion with typical fluxes $F\gsim$100 $\mu$Jy at 8.4 GHz.  Through a Population SYnthesis Code coupled with the standard afterglow Hydrodynamical Emission model (PSYCHE) we reproduce the radio flux distribution of the radio sub--sample. Based on these results we study the detectability in the time/frequency domain of the entire long GRB population by present and future radio facilities. We find that the GRBs that typically trigger \sw\ can be detected at 8.4 GHz by JVLA within few days with modest exposures even at high redshifts. The final SKA can potentially observe the whole GRB population provided that there will be a dedicated GRB $\gamma$--ray detector more sensitive than \sw. For a sizable fraction (50\%) of these bursts, SKA will allow us to perform radio--calorimetry, after the trans--relativistic transition (occurring $\sim$100 d), providing an estimate of the true (collimation corrected) energetics of GRBs. 
\end{abstract}

\section{Introduction}

Although the afterglows of Gamma Ray Bursts (GRBs) have been studied for 16 years, their complexity and variety has been explored only recently thanks to the rapid location and repointing of the counterparts of GRBs by the \sw\ satellite (Gehrels et al. 2004). The study of  GRBs at radio frequencies is still poor compared to the wealth of information available at  other wavelengths (X-ray and optical/NIR).  The most recent extensive study of the radio afterglows of GRBs (Chandra \& Frail 2012 - CF12 hereafter) collected 304 GRBs (from January 1997 to April 2011) that were observed (between $\sim$0.1 and $\sim$300 days) in the radio bands (at different frequencies between 600MHz and 660GHz - the 8.4 GHz frequency being the most sampled one). The detections have typical flux densities of 150-200 $\mu$Jy and the 3$\sigma$ upper limits are at the level of 100-150$\mu$Jy. The current detection rate (fraction of GRBs detected with respect to observed) in the radio band is $\sim$30\% (CF12), mainly limited by the sensitivity of the radio facilities available until the recent Karl G. Jansky Very Large Array (JVLA). JVLA, which can reach a sensitivity limit of few tens of $\mu$Jy for continuous observations at $\sim$8.4 GHz, opens the path towards the forthcoming Square Kilometer Array (SKA). 

Despite their paucity, observations of GRBs in the radio band have proved some fundamental properties of these sources: (1) radio scintillation (GRB 970508 -- Frail et al. 1997) showed that the outflow is relativistic; (2) the late time ($\sim$100 d) flattening of the light curve (GRB 980703 and 000418 -- Frail et al. 2004) has been interpreted as the jet becoming non--relativistic; (3) late time (100--450 d after the burst) radio calorimetry  (GRB 030329 -- Frail et al. 2005, van der Horst et al. 2005, 2008; GRB 970508, 980703 -- Berger, Kulkarni \& Frail  2004) provided an estimate of the total kinetic power of the jet; (4) the afterglow spectral energy distribution, obtained by combining optical and radio data, supports the synchrotron origin of the radiation emitted during the afterglow phase; (5)  radio monitoring of local SN Ibc   put some constraints on the GRB/SN association (e.g. Berger et al. 2003) and on the burst beaming factor (Soderberg et al. 2006); (6) radio discovery of SN2009bb--like objects (Soderberg et al. 2009) can shed light on the transition between normal SNe and engine--driven like GRB; (7) the detection of  a few GRB hosts (Stanway et al. 2010; Hatsukade et al. 2012; Zauderer et al. 2012) in the radio band provides an estimate of their unobscured star formation rate.  

 The radio flux distribution of GRBs at a given frequency and time depends on the specific dynamical evolution of the fireball, on the emission mechanism (synchrotron) and on the burst population properties (e.g. the redshift distribution, the energetics and the opening angles).  Predictions of the radio properties of GRBs as a population is of fundamental importance in preparation for the Square Kilometer Array (SKA).

In the lead-up to the SKA, several next--generation radio telescopes are being constructed around the world (see e.g. Norris et al. 2012), including the two official SKA precursors: ASKAP (Australian SKA Pathfinder) and MeerKAT. ASKAP (Johnston et al. 2007, 2008; Deboer et al. 2009) is a new radio interferometer being built in Western Australia, and is expected to start operations in 2014 with a limited number of antennas. In its final design, ASKAP will consist of 36 12-m antennas distributed over a region of 6 km in diameter. Each antenna will be equipped with a PAF (Phased Array Feed), operating in a frequency band of $0.7-1.8$ GHz. PAF technology allows very wide field of views ($\sim 30$ deg$^2$ at 1.4 GHz), at the expenses of sensitivity. This makes ASKAP best suited for wide all--sky surveys, rather than pointed deep integrations.  MeerKAT (Jonas 2009) is the South African SKA precursor telescope. It will be constructed in two phases. MeerKAT1 (2016) will consist of  64 13-m antennas distributed over a region of 8 km in diameter. It will be equipped with single-pixel receivers, operating in a limited frequency range ($0.9-1.7$ GHz). The frequency range will be significantly extended in phase 2 (MeerKAT2, 2018), with operational bands $0.6-1.7$ GHz and $8-14.5$ GHz. For MeerKAT2 the telescope baselines will be possibly increased up to 16 km, allowing better spatial resolution and consequently deeper integrations.       

With a collecting area of about a square kilometre, the SKA will be far superior in sensitivity and observing speed to all current radio facilities\footnote{For more details on SKA specifications we refer to the official documentations at the SKA web site:
www.skatelescope.org}. The SKA is also designed to be built in two phases (SKA1 and SKA2), expected to become fully operational by 2020 and 2025, respectively. In each phase SKA will consist of three elements: one operating at low frequency, one at intermediate/high frequency, and one optimized for surveys. In the following we focus on SKA performance at mid/high frequencies, which are most relevant for this study. The mid/high frequency band covered by SKA1 is $0.45-3$ GHz, to be extended to 10 GHz in phase 2 (SKA2). SKA1 will consist of 250 15-m antennas distributed in a region of $\sim 100-200$ km in diameter. SKA1 is designed to provide factor $\sim 5$ better sensitivities than MeerKAT. SKA2 will consist of $\sim 2500$ antennas with an increase in sensitivity of another factor 10. 

In this paper (\S2) we collect all the available radio fluxes (detections and upper limits) of a complete sample (BAT6 - Salvaterra et al. 2012) of bright \sw\ long bursts. We combine (\S3) a population synthesis code (Ghirlanda et al. 2013) with a hydrodynamical-radiative code (van Eerten, van der Horst \& MacFadyen 2012) to compute the afterglow radio emission of the population of GRBs. Our model (\S4) reproduces the observed radio flux distribution of the complete \sw\ sample and allows us to make predictions for the radio detection rate by current and future radio telescopes from JVLA to the SKA. We discuss our findings in \S5.  
A standard cosmological flat Universe with $h=\Omega_{\Lambda}=0.7$  is assumed.

\section{The BAT6 complete sample: radio properties}
\begin{table}
\caption{Radio flux at 8.4 GHz of the GRBs of the BAT6 sample. (*) The flux has been extrapolated to 8.4 GHz using the flux at the  frequency (shown in parenthesis) assuming that the radio emission at the corresponding time was in the self--absorbed regime. (**) The flux was extrapolated backward (with respect to the reported time of the  observation) to 2 days adopting a typical $t^{1/2}$ scaling (CF12). Upper limits are at the 3$\sigma$ level of confidence. References: 
(1) Berger et al. 2005 ; 
(2) Chandra \& Frail 2012; 
(3) Frail \& Chandra private communication; 
(4) Chandra \& Frail 2006; 
(5) Chandra \& Frail 2006a;
(6) Chandra \& Frail 2007; 
(7) Cenko et al. 2010; 
(8) Chandra \& Frail 2008; 
(9) Soderberg \& Frail 2008; 
(10) Chandra \& Frail 2008a; 
(11) Chandra \& Frail 2009; 
(12) Chandra \& Frail 2009a; 
(13) Chandra \& Frail 2009b; 
(14) Frail \& Chandra 2009c; 
(15) Greiner et al. 2013; 
(16) Zauderer et al 2011; 
(17) Frail D. A. et al. 2011; 
(18) Pooley 2005; 
(19) van der Horst, 2006; 
(20) Frail et al., 2006; 
(21) Chandra \& Frail, 2006b; 
(22) Chandra \& Frail, 2006c; 
(23) van der Horst, 2006a; 
(24) Chandra \& Frail, 2007a; 
(25) Chandra \& Frail, 2007b; 
(26) Chandra \& Frail, 2008b; 
(27) Chandra \& Frail, 2008c; 
(28) Chandra \& Frail, 2008d; 
(29) van der Horst, 2008a; 
(30) Moin et al., 2008; 
(31) Chandra \& Frail, 2008e; 
(32) Chandra \& Frail, 2008f; 
(33) Chandra \& Frail, 2009c; 
(34) Chandra \& Frail, 2009d; 
(35) Chandra \& Frail, 2009e.
}
\begin{tabular}{lcccc}
\hline
GRB        &	t 	& 		F 				 		& 		Instr.(GHz)	 & 		Ref \\
		  &  [days]	&		[mJy]				&			&						\\
\hline
050401 	  & 5.69	&      0.122$\pm$0.033 	 			& 	       VLA  	 & 		1     \\
050416A  & 5.58	&      0.260$\pm$0.055 	 			&  		VLA  	 &		1     \\
050525A**  &13.5  &   	 0.063						&      	VLA  	 &		2	\\
050922C & 1.34		&	0.140$\pm$0.042			&		VLA		 &		19	\\		
061121 	  &1.16 	&	 0.304$\pm$0.048    	 		&		VLA  	 &	 	4	\\
061222A  &1.00	&	 0.285$\pm$0.068			&		VLA		 &		5	\\
071020 	  &2.18 	&	 0.186$\pm$0.026   	 			&		VLA  	 &		6 	\\
080319B  &2.30 	&	 0.232$\pm$0.042   		 		&		VLA  	 &		7	\\	
080413B  &1.22 	&	 0.086$\pm$0.036   			&		VLA  	 &		8	\\
081007 	   &1.97 	&	 0.320$\pm$0.030   			&		VLA 	 &	 	9	\\
081221 	   &5.4  	&	 0.097$\pm$0.038	 			&		VLA		 &  		10	\\ 
090424	   &1.5   &  	 0.673$\pm$0.039   			& 		VLA  	&		11	\\ 
090715B   &4.26 &   	 0.231$\pm$0.047   			&		VLA  	&		12   \\
090812	    &1.12 & 	 0.104$\pm$0.043   			&		VLA  	&		13 	\\
091020     &3.7 	&	 0.230$\pm$0.042   			&		VLA  	&		14	\\ 
100621A*  &4.0		&	 0.130$\pm$0.024			&		ATCA(9.0)	&		15	\\
110205A* &1.2	&	 0.026$\pm$0.002			&		JVLA(22)	&		16   \\
110503A* &0.3	&	 0.031$\pm$0.003							&		JVLA(19)	&		17   \\
\hline
050802A* &	0.36&	 $\le$0.094					&		Ryle(15)	&		18	\\
060206*	 &     2.0	&       $\le$0.193					&		WSRT(4.9)	&		19	\\	
060210A   & 4.0	&	$\le$0.072					&		VLA		&		20	\\
060908A   &	2.0	&	$\le$0.075					&		VLA		&		21	\\
060912A   &	2.0	&	$\le$0.097						&		VLA		&		22	\\
061007	   &	1.0	&	$\le$0.141					&		ATCA	&		23	\\
070306	   &	1.52&	$\le$0.090					&		VLA		&		24	\\
071112C  & 3.56	&	$\le$0.114					&		VLA		&		25	\\
080430	   &0.34	&	$\le$0.138				&		VLA		&		26	\\
080603B  &1.34	&	$\le$0.123					&		VLA		&		27	\\
080721	   &2.72	&	$\le$0.144				&		VLA		&		28	\\
080804	   &0.9	&	$\le$0.189   					&		ATCA	&		29	\\
081121*	   &3.5	&	$\le$1.839					&		ATCA(4.8)	&		30	\\
081203A   &0.81 &	$\le$0.162   					&		VLA		&		31	\\
081222	   &0.97	&	$\le$0.159					&		VLA		&		32	\\
090102     &1.23	&	$\le$0.147   				&		VLA		&		33	\\
090709A   &2.1	&	$\le$0.105						&		VLA		&		34	\\
091127	   &2.24	&	$\le$0.300						&		VLA		&		35	\\
\hline
\end{tabular}
\label{tab1}
\end{table}

Complete samples are the basis for any statistical study of astrophysical sources. Recently, Salvaterra et al. (2012) assembled a well selected complete sample of 58 long GRBs (BAT6). GRBs are selected on the basis of favorable observing conditions and by their brightness in the {\it Swift}/BAT 15--150 keV band, i.e. with 1-s peak fluxes $P\ge$ 2.6 ph s$^{-1}$ cm$^{-2}$. This value is much larger than the detection limit of \sw/BAT ensuring that the sample is  complete with respect to this flux limit. This means that all the GRBs with a peak flux $\ge$ 2.6 ph s$^{-1}$ cm$^{-2}$ would have been firmly detected by \sw/BAT. The BAT6 sample has also a large completeness in redshift: 95\% of GRBs in the sample have a known redshift (see Covino et al. 2013 for the last update of redshifts). Therefore, the BAT6 sample is one of the best datasets to explore the properties of the long bright GRB population and their redshift evolution. Moreover, BAT6 represents a critical test for long GRB population synthesis models (Ghirlanda et al. 2013) or to study the impact of instrumental biases on the intrinsic GRB spectral--energy correlations (Ghirlanda et al. 2012a) and to test possible new theoretical models for the central engine (Bernardini et al. 2013).

Here, we collect all the radio observations for the bursts in our complete sample. 38 out of 58 GRBs (68\%) have been observed at radio frequencies, mainly by VLA and ATCA (see Table 1).
We check that the radio observed sub-sample preserves the properties of the original one. A KS test provides a probability always larger than 0.95 that the distribution of redshifts, prompt emission properties (Nava et al. 2012), X-ray equivalent hydrogen column densities (Campana et al. 2012), afterglow X-ray luminosity (D'Avanzo et al. 2012) and dust extinction (Covino et al. 2013) are drawn from the same parent population. Also the fraction of dark bursts in the radio observed sub-sample is consistent with the one of the original sample (Melandri et al. 2012). This is the first GRB sample selected by the \sw\ trigger in the $\gamma$--ray band and it represents an unbiased sample to study the radio properties of GRB afterglows. The larger sample of CF12 suffers from the fact that its mostly restricted to extensively followed GRBs at optical and X--ray frequencies or to interesting cases (e.g. GRB/SN, high redshift bursts).

For half of the BAT6 bursts observed in the radio band, at least one detection at radio frequencies is reported. This percentage is larger than the detection rate ($\sim$30\% CF12) for pre-{\it Swift} or {\it Swift} radio samples. This fact suggests that bright GRBs in the {\it Swift}/BAT band tend to be bright also in the radio bands (as will be shown also in \S4). Radio fluxes and upper limits for our well selected sub-sample are reported in Table 1. Radio observations are generally very sparse. However the large majority (76\%) of bursts has been observed at 8.4 GHz and within 1-5 days from the trigger. We homogenize the remaining bursts by extrapolating the observed flux or limits to the 8.4 GHz band (labelled with single asterisk in Table \ref{tab1}) and to a common observer frame time of $\sim 2$ days (double asterisks in Table \ref{tab1}) from the explosion. We note that this was not possible for two bursts in our sample, namely 060614 and 090201, for which only late time upper limits are available. Extrapolation of these two upper limits to early time is hampered by their peak times being unknown. We therefore exclude these two bursts from our analysis.

We find that the median radio flux at $\sim 2$ days from the trigger for the $\gamma$-ray bright bursts is $\approx 100$ $\mu$Jy, with only a small fraction of bursts in our sample (3-6\%) brighter than 500 $\mu$Jy in the 8.4 GHz band.

\section{Population Synthesis Code and Hydro-Emission model (PSYCHE)}

We have  combined a Population SYnthesis Code (Ghirlanda et al. 2013 - G13 hereafter) with a Hydrodynamical and Emission model (van Eerten 2012 - VE12 hereafter - van Eerten \& MacFadyen 2012a, 2012b) obtaining ``PSYCHE" that works in two steps:
\begin{enumerate}
\item[(1)] the first step consists in generating a population of GRBs distributed in the Universe that reproduces a set of GRB properties in the hard X--ray and $\gamma$--ray band. The population synthesis parameters are adjusted in order to reproduce the rate, peak flux and fluence distributions of the populations of GRBs detected by  \sw, \cgro\ and \fe\ and the intrinsic \ama\ correlation (between the rest frame $\nu F_{\nu}$ peak energy and the isotropic equivalent energy, respectively) of the BAT6 sample. This Montecarlo code was presented in G13 (see also Ghisellini et al. 2013 for a summary) and the main result is a synthesized population of GRBs each one with an assigned redshift $z$, isotropic equivalent energy $E_{\rm iso}$, jet opening angle \th, initial bulk Lorentz factor $\Gamma_0$ and viewing angle $\theta_{\rm view}$ with respect to the observer line of sight. Therefore, each simulated GRB is described by a set of parameters: [$z$, $E_{\rm iso}$, $\Gamma_0$, \th, \thv];
\item[(2)] the second step is to use the output parameters of the population code  to calculate the flux at any given time and frequency. To this aim we adopted the code of  VE12 that describes the afterglow emission of an adiabatic fireball expanding into a constant density interstellar medium. 
This code accounts for the full hydrodynamical fireball evolution, i.e. from the early relativistic to the late time non--relativistic phase.  
However, the code includes only the emission from the forward shock. The reverse shock emission, which might be important at early times (e.g. $\sim$1 day), is not included. 
The emission mechanism is Synchrotron and self--absorption, particularly important for the predictions of the radio emission, is accounted for. This code computes the emissivity also accounting for the orientation of the GRB jet with respect to the observer line of sight (i.e. \thv). The input parameters necessary to compute the flux of each GRB at a given frequency and time are: the isotropic equivalent kinetic energy $E_{\rm k,iso}$ driving the afterglow expansion into the interstellar medium (ISM), its density $n$, the jet opening angle \th\ and  the viewing angle \thv. \th\ and \thv\  are the result of the population synthesis code of G13 (step 1 above), the kinetic energy is derived assuming that the radiative efficiency has a typical value $\eta$=20\% for all bursts so that $E_{\rm k,iso}=E_{\rm iso}/\eta$.  Synchrotron emission is regulated by the microphysical parameters at the shock:  the electron power law energy index $p$, the fraction of electrons accelerated at the shock front (which we assume to be 1)  and the fraction of the shock energy shared between  electrons and  magnetic field $\epsilon_{e}$ and $\epsilon_{B}$, respectively. 
\end{enumerate}

The scope of this paper is to predict the {\it radio flux distribution} of the population of GRBs that can trigger present and future $\gamma$--ray burst detectors: these are the bursts pointing towards the Earth. As explained in G13, these  bursts have a viewing angle \thv$\le$\th\ or  \thv$\le$arcsin(1/\G) (where 1/\G\ represents the angle within which the prompt radiation is beamed).

\section{Results}
\begin{figure}
\hskip -0.3truecm
\includegraphics[width=8.5cm,trim=20 10 20 20,clip=true]{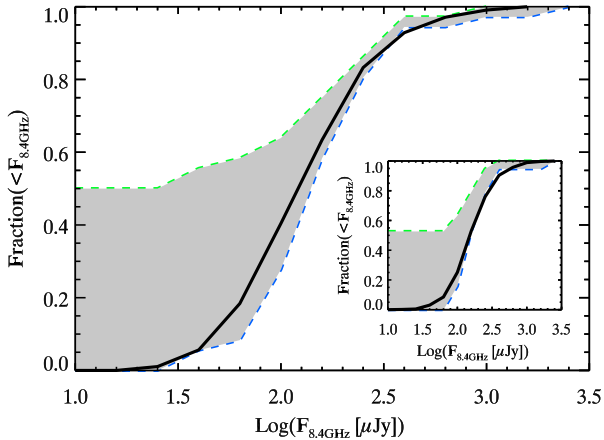}
\caption{Cumulative normalized flux distribution (at 8.4 GHz) of radio observations for the GRBs of the BAT6 sample (Table \ref{tab1}).  
The y--axis shows the percentage of GRBs with flux smaller than the corresponding flux value on the x--axis.
The shaded region is delimited by the cumulative distribution (dashed [green] upper boundary) obtained assuming 
that all the bursts with upper limits have no radio emission and by the cumulative distribution (dashed [blue] lower limit) obtained assuming that all the upper limits are detections. The solid (black) line shows the model that reproduces the BAT6 radio flux distribution.  
Inset: same as main plot but considering only the GRBs of the BAT6 sample (Table \ref{tab1}) with detections and upper limits at $t>3$ days. Also the  model (solid line) is recomputed at $t>3$ days with the same parameters adopted for the model shown in the main plot. 
}
\label{fg1}
\end{figure}

Our {\it observational constrain} is the flux distribution of the bright bursts of the BAT6 \sw\ sample reported in Table \ref{tab1}. This is shown by the grey shaded region in Fig.~\ref{fg1}  (main plot). The upper limit curve of this region is obtained by combining the radio detections (first part of Table \ref{tab1}) and bursts with upper limits (second part of Table \ref{tab1}) have no radio emission. The lower boundary is obtained considering  the detections together with the upper limits (the latter assumed as detections). Any flux distribution resulting from PSYCHE, when a set of parameters [$n$, $p$, $\epsilon_e$, $\epsilon_B$] is assigned, should lie in the grey shaded region.

Given the lack of knowledge of the distributions of the microphysical parameters of GRBs and the strong degeneracy among them, we considered the simplest assumptions that can reproduce the flux distribution of Fig.~\ref{fg1}. The scope of the present paper is not a parametric study of the microphysical parameters of the shock. 
We can reproduce the flux distribution of the BAT6 sample fixing $p=2.5$, $\epsilon_{e}=0.02$ and $\epsilon_{B}=0.008$ and assuming that  $n$ has a uniform distribution between 0.1 and 30 cm$^{-3}$. The assumed values of $p$, $\epsilon_{e}$ and $\epsilon_B$ are consistent with the typical values derived from the afterglow modeling of GRBs (Panaitescu \& Kumar 2000; Ghisellini et al. 2009) and  with what derived from recent first-principles simulations of particle acceleration in relativistic shocks (Sironi, Spitkovsky \& Arons 2013; Martins et al., 2009; Haugbolle  2011). In particular, Sironi et al. (2013) find that the magnetic energy fraction reaches $\epsilon_B\sim$0.1 in the shock transition layer, but its average value in the region where particles are accelerated (and so, where the synchrotron emission will be produced) is smaller, around $\epsilon_B\sim$0.003--0.01. Regarding $\epsilon_e$, Sironi et al. (2013) report that the fraction of shock energy given to non-thermal electrons is commonly $\epsilon_e\sim$0.03, being the product of the fraction of shock energy given to electrons (thermal and non-thermal), which is typically 30\%, and the fraction of electron energy contained in the non-thermal tail, of order 10\%.  We anticipate here that with these values the PSYCHE code can also reproduce the optical flux distribution of the BAT6 sample and these results will be presented in a forthcoming paper (Ghirlanda et al. 2013). 

The PSYCHE model (solid line) shown in Fig.~\ref{fg1} is obtained considering, within the population of synthesized bursts (with microphysical parameters set as described in the previous section), those with a peak flux (in the 15--150 keV energy range) $P\ge$ 2.6 ph cm$^{-2}$ s$^{-1}$. This flux limit corresponds to the selection criterion of the BAT6 sample which is our observational constrain.  
The PSYCHE model shown in Fig.~\ref{fg1} is obtained by simulating, for each burst, the radio flux at a random epoch between 1 and 5 days, corresponding to the epochs of the radio observations of the BAT6 sample (Col.1 in Tab. \ref{tab1}). However, at early times ($\sim$1 day) it is possible that the radio flux is dominated by the reverse shock component (e.g. Kulkarni et al. 1999; Gomboc et al. 2009) which is not modeled in our code. Instead, the later time emission (at $t\ge3$ days) is more likely dominated by the forward shock component. As a consistency check we verified (inset of Fig.\ref{fg1}) that the model that reproduces all the radio observations (reported in Tab.\ref{tab1}) is also able to reproduce the radio observations (detections and upper limits) at late times (i.e. $\ge$3 days).

\subsection{Radio detection rate}

The past observation strategy of GRB afterglows in the radio band is somewhat inherited from the optical experience and it mainly consisted in performing early time observations, within few days since the burst detection, with the main scope of following the afterglow also at low frequencies and with the aim of performing multi wavelength studies by combining optical/NIR and radio data. There are, however, few bursts that were observed up to extremely late times (Frail et al. 1997, 2004; Berger et al. 2004; van der Horst 2005, 2008).  Observations at $t>2$ days might more successfully detect the afterglow around the peak of the light curve (when it is brighter - see also \S4.2).  
Due to the fact that our code does not account for the possible contribution of the reverse shock component at early times, we have computed the detection rate 
of  present and future radio telescopes for a typical observation at about  5 days  
when the forward shock should dominate the emission. The detection rates are given considering a band approximately centered around 8.4 GHz, this is indeed the typical sampling frequency of presently available radio GRB observations, performed in most cases, with the VLA (see Table \ref{tab1}). 

\begin{figure}
\includegraphics[width=8.5cm,trim=20 10 20 20,clip=true]{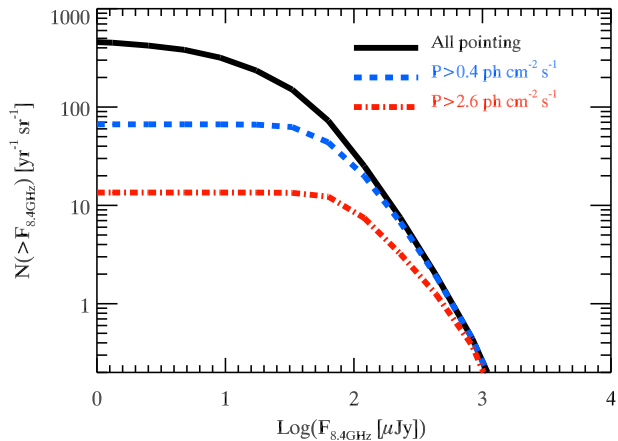}
\caption{Cumulative flux distribution 
(i.e. number of objects per year and unit solid angle with flux larger than the 
corresponding value on the x--axis)
at 8.4 GHz for all simulated GRBs pointing to the Earth (black solid line) 
and for the \sw\ subsample (i.e. with peak flux  $P\ge$ 0.4 ph cm$^{-2}$ s$^{-1}$
- dashed blue line) and the \sw\ subsample of bright bursts (i.e. 
$P\ge$ 2.6 ph cm$^{-2}$ s$^{-1}$, dot--dashed red line). 
}
\label{fg2}
\end{figure}

Fig.~\ref{fg2} shows the cumulative rate distributions of the radio flux at 8.4 GHz computed at  5 days in the observer frame. The PSYCHE sample of synthesized bright bursts (i.e. $P\ge$ 2.6 ph cm$^{-2}$ s$^{-1}$ corresponding to the BAT6 sample)  is shown by the (red) dot--dashed line while the  (blue) dashed line shows the distribution for bursts with $P\ge$ 0.4 ph cm$^{-2}$ s$^{-1}$. This flux cut corresponds to the detection limit of the Burst Alert Telescope (BAT) onboard \sw. The total sample of simulated bursts pointing towards the Earth is shown by the (black) solid line. 

As can be seen in Fig.~\ref{fg2} most of the bursts can be detected at 8.4 GHz with early time observations if a sensitivity of a few $\mu$Jy is reached. This low flux limit can now be reached  with the JVLA and will be easily reached also by SKA precursors, like MeerKAT, and SKA itself. 
Assuming a typical bandwidth (reported in parenthesis in Col. 2 of Table \ref{tab2}) and an integration time of 30 minutes for continuum observations we derive the detection limits (at 3$\sigma$ significance) reported in Col. 3 of Table \ref{tab2}. The sensitivies reported in Table \ref{tab2} can be rescaled approximately for the assumed bandwidth and for the exposure time as $(\Delta\nu\, t_{exp})^{1/2}$. Col. 4 of Table~\ref{tab2} gives the detection rate in the radio band $\mathcal{R}_{b}$ of the bursts that are detected by \sw--BAT (i.e. with peak flux $P\ge$ 0.4 ph cm$^{-2}$ s$^{-1}$ - dashed blue line in Fig.~\ref{fg2}).  Similar detection rates are obtained for JVLA and ``MeerKAT2" (i.e. the ``phase 2" implementation of the South Africa SKA precursor). Access to the complete population of GRBs (solid black line in Fig.~\ref{fg2}), i.e $\mathcal{R}$=478 GRBs yr$^{-1}$ sr$^{-1}$, will be possible with ``SKA2" (i.e. the ``phase 2" - final implementation of SKA). 

Considering high redshift GRBs\footnote{High redshift GRBs simulated by PSYCHE do not include Pop--III events as the $z \ge$ 6 population is thought to be dominated by Pop--II GRBs (Campisi et al. 2011). }, JVLA or MeerKAT2 can see all the $z\ge$ 6 GRBs detected by \sw, while SKA2 will asses the entire high redshift GRB population.

Note that if we consider the population of GRBs presently detected by BAT onboard \sw, the detection rates of JVLA, MeerKAT2 and the final SKA are comparable ( 67 GRBs yr$^{-1}$ sr$^{-1}$). These rates would increase, already with the JVLA, only if a  $\gamma$--ray detector more sensitive than \sw-BAT be available in the near future.  

\begin{table}
\caption{Detection rate for observations at  5 d at 8.4GHz with present and future radio facilities. $\mathcal{R}$ ($\mathcal{R_{\rm b}}$) are the rate of GRBs pointing to the Earth detectable in the radio band of the \sw\ entire (bright) population. }
\begin{center}
\begin{tabular}{lcccc}
\hline
Tel.       	   &	$\nu_{\rm obs}$($\Delta\nu$) 	& 	$F_{\rm lim}$ 	& 	$\mathcal{R_{\rm b}}$	 & 	$\mathcal{R}$	 \\
	           &  [GHz] 	&	[$\mu$Jy]		&				 	 [\#/yr sr]  &[\#/yr sr]		 				\\
\hline
JVLA	         &  8.4(1.024)			& 	16			&	66					&	246			\\
MeerKAT2    &  8.4(2.0)				&	16			&	66					&	246		\\	
SKA2	        & 8.4(2.0)				&	0.3			&	67					&	478			\\
\hline
\end{tabular}\end{center}
\label{tab2}
\end{table}

\begin{figure}
\includegraphics[width=8.5cm,trim=20 10 20 20,clip=true]{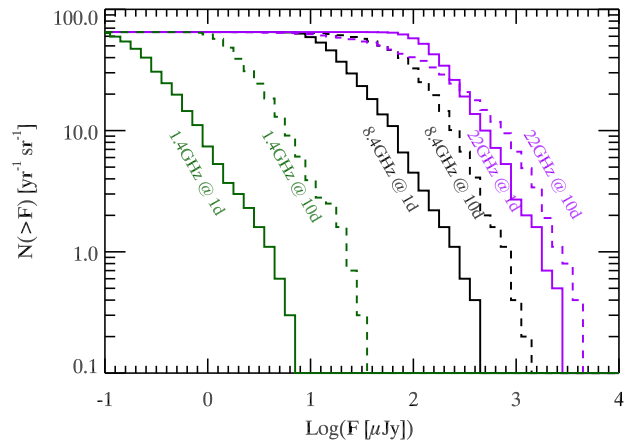}
\caption{Cumulative flux distributions for the bursts pointing to the Earth but limited to those with peak flux $P\ge$0.4 phot cm$^{-2}$ s$^{-1}$ (representative of the flux limit of the \sw\ burst population -- i.e. blue triangles and red squares in Fig.~\ref{fg4}). Two sampling times are shown: 1 day (solid lines) and 10 days (dashed lines). Three sampling frequencies are shown: 1.4 GHz (green lines), 8.4 GHz (black lines) and 22 GHz (purple lines).
}
\label{fg3}
\end{figure}

Early time radio observations are fundamental for the multi wavelength study of GRB afterglows and to detect radio scintillation such as in GRB 970508 (Frail et al. 1997). The radio detection rate  increases, for any observing frequency in the typical radio band, if the observations are performed at relatively later times. Figure \ref{fg3} shows the flux at 10d (dashed line) which is systematically larger than at 1d (solid line).  
As noted before, our modeling does not include the reverse shock  which could  contribute to the early time radio 
emission.
The bursts shown here are the \sw\ events, i.e. those that can trigger BAT. From Fig.~\ref{fg3} it is possible to derive the detection rate at the typical radio frequencies (1.4, 8.4 and 22 GHz) for early and intermediate time observations.


\subsection{Peak radio emission}

The radio flux is initially self--absorbed until, due to the expansion, the emission region becomes optically thin. Practically, this corresponds to the passage of the synchrotron self--absorption frequency across the observing frequency. This transition, happening first at higher frequencies, corresponds to the peak of the radio afterglow light curve. Through PSYCHE we have computed the light curves of the synthesized GRBs up to very late time and searched for the peak time $t_{peak}$ (in the observer frame) and the corresponding peak flux $F_{peak}$ at three observing frequencies. 
In Fig.~\ref{fg4} we show the distribution of the peak fluxes versus the observed peak times for the entire population of GRBs  
pointing to the Earth: bursts with peak flux $P<0.4$ phot cm$^{-2}$ s$^{-1}$ are shown by the black circles,  the \sw\ bursts with $0.4\le P < 2.6$ phot cm$^{-2}$ s$^{-1}$ are shown by the blue triangles and the \sw\ bursts with $P\ge2.6$ phot cm$^{-2}$ s$^{-1}$ are shown by the red squares. 
 The peak is brighter and reached at earlier times at higher frequencies. High energetic bursts populate the upper right part of the $F_{peak}-t_{peak}$ plane shown in Fig.~\ref{fg4}. A small fraction of GRBs ($\sim$1.5\% of \sw\ bursts and 5.3\% of \sw\ bright bursts) should be $>$1 mJy at the peak (e.g. at 8.4 GHz in Fig. \ref{fg4} middle panel). Such events are brighter than 100$\mu$Jy at $\simeq$100 days as indeed observed in some cases (Shivvers \& Berger  2011 and references therein). However, this fraction is likely to be underestimated in our simulations because small statistics dominate the high(low) $\gamma$--ray luminosity(redshift) distribution. On the other hand,  the ease to follow up these events make them over--represented in current radio samples. Moreover, such events are likely to have extreme microphysical parameters which are not represented in our simplified assumptions.

Considering for example the Australian SKA precursor ASKAP, for a maximum bandwidth of 300 MHz centered at 1.4 GHz it should reach 157$\mu$Jy (3$\sigma$), so that it will detect a small fraction of GRB afterglows (top panel of Fig.~\ref{fg4}). MeerKAT1 at 1.4 GHz should reach a $3\sigma$ flux limit of 23$\mu$Jy thus being able to detect $\sim$80\%  of the \sw\ bursts provided that the observations will be performed at $t\ge$10 days. Fig.~\ref{fg4} shows that higher frequency observations (e.g. 8.4 GHz) can detect the peak at relatively earlier times ($\sim$1-10 days).


\subsection{Population III GRBs}

Several recent works have shown that metal free, very massive first stars (so called Pop--III stars) can produce a GRB event (Meszaros \& Rees 2010; Kommisarov \& Barkov 2010; Suwa \& Ioka 2011). In particular Toma, Sakamoto \& Meszaros (2011) suggest that Pop--III bursts could have an equivalent isotropic energy $E_{\rm iso}\sim 10^{57}$ erg and an extremely long duration $t\sim10^4$ seconds (set by the timescale of accretion of the thick cold disk and possibly stretched by fallback). They also show that these events are easily detectable in the radio band despite their extreme redshifts. 

\begin{figure}
\includegraphics[width=9cm,trim=40 0 60 0,clip=true]{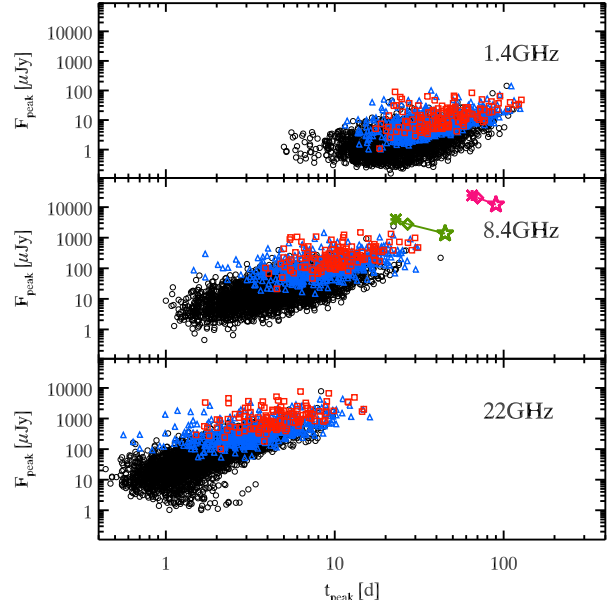}
\caption{Peak flux versus time of the peak (observer frame) of the light curve at three frequencies (as labelled). All bursts pointing to the Earth are represented:  black circles are the bursts with peak flux $P < 0.4$ phot cm$^{-2}$ s$^{-1}$,  blue triangles are the 
bursts with peak flux $0.4 \le P <2.6$ phot cm$^{-2}$ s$^{-1}$ and red squares are the bursts with peak flux $P\ge$2.6 phot cm$^{-2}$ s$^{-1}$ (representative of the \sw\ bright population, i.e. the BAT6 sample).  Two cases for a typical GRB (\th\ = 0.1 radiants, \thv\ = 0 radiants, $n=1$ cm$^{-3}$ and the same microphysical parameters adopted in \S3) are shown in the middle panel for two values of the isotropic equivalent kinetic energy: $E_{k,iso}=5\times10^{54}$ erg and  $E_{k,iso}=5\times10^{55}$ (green and magenta symbols) for three redshifts, $z=5, 10, 20$ (asterisk, diamond and star, respectively). 
}
\label{fg4}
\end{figure}

In Fig.~\ref{fg4} we show (middle panel) the position of a GRB with isotropic equivalent energy $E_{\rm iso}=10^{54}$ erg (i.e. kinetic energy $5\times10^{54}$ erg - green symbols) and $E_{\rm iso}=10^{55}$ erg (i.e. kinetic energy $5\times10^{55}$ erg -  magenta symbols). The latter is representative of Pop--III events. We considered three possible redshifts: $z=5$, 10 and 20 (asterisk, diamond and star, respectively). Pop--III events are easily detected with peak fluxes of 10 mJy and they should peak at much larger times ($\sim$60-100 days) with respect to the normal GRB population. These numbers depend only slightly on the redshift because the time dilation is counterbalanced by the negative k--correction.  

In the absence of any observational tool to identify candidate Pop--III events, we suggest that the detection of a radio afterglow peaking at relatively high flux levels at late times could be a signature of a high redshift energetic (Pop--III) event. This is particularly true for $z\gsim19$ events that are missed even in the K band due to IGM absorption. 

 We note that the position in the $t_{peak}-F_{peak}$ plane of GRB~090423 at $z=8.2$ (Salvaterra et al. 2009; Tanvir et al. 2009) as derived by radio measures (Chandra et al. 2009) makes it consistent with the tail of the normal (Pop--II) GRB population. This further supports the idea that this burst in spite of its extreme redshift was not powered by the explosion of a massive Pop--III star.
 
\subsection{Trans--relativistic transition and radio calorimetry}

The standard afterglow model predicts that the outflow should become non--relativistic (NR) at $t_{NR}$ typically around 100 d (Livio \& Waxman 2000). 

\begin{figure}
\includegraphics[width=8.5cm,trim=0 0 0 0,clip=true]{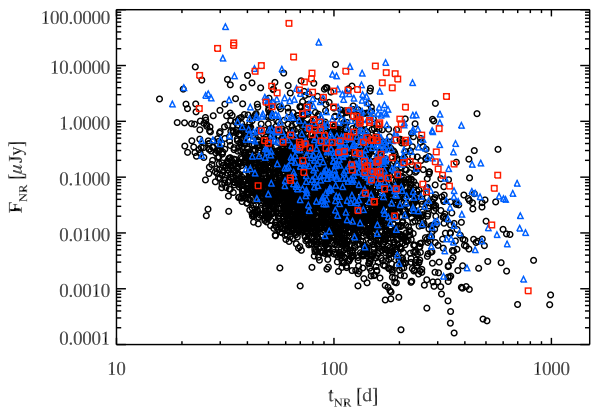}
\caption{Flux at the transition to the non--relativistic phase versus the time of this transition. Symbols as in Fig.~\ref{fg4}
}
\label{fg5}
\end{figure}

Figure \ref{fg5} shows the flux $F_{NR}$ at the time of the trans--relativistic transition $t_{NR}$, derived from the formalism of Livio \& Waxman (2000), for the simulated GRB population. With the present and near future facilities (JVLA and MeerKAT) the study of the trans--relativistic transition is possible only for a small fraction of GRBs. Late time observations across the trans--relativistic transition could instead be routinely done with the final SKA, allowing  population studies. With the final SKA sensitivity reported in Table~\ref{tab2} we expect that $\sim$50\% of the \sw\ bursts will be detected at the trans--relativistic transition. This number increases with longer exposures. 

The major issue for late time observations (across $t_{NR}$) might be represented by the possible contamination from the host galaxy emission. Although present studies of the GRB hosts at radio frequencies (e.g. Mickalowski et al. 2012) 
are based in most cases on upper limits, we can roughly estimate the flux of the host (at the redshift of each synthesized burst) at 8.4 GHz, assuming the standard equation of Yun \& Carilli (2002) and assuming a standard host spectrum (with $\beta=-0.6$ - e.g. Berger et al. 2003). For the median value of the long GRB host star formation rate of 2.5 $M_{\odot}$ yr$^{-1}$  (Savaglio, Glazebrook \& Le Borgne 2009) we estimate that 80\% of the \sw\ bursts are brighter than their hosts at $t_{NR}$ and 97\% of these can be detected by SKA with a 30 min exposure. Even if the  star formation rate would be larger by a factor of 10 the latter fraction would still be sufficiently large, $\sim$40\%, to allow radio calorimetry studies for a sizable fraction of the bursts. This will allow us to study the true energetics for a large statistical sample of long GRBs.

\section{Discussion and Conclusions}

We  collected the available radio observations for the BAT6 sample. This is a complete sample of bright \sw\ GRBs (Salvaterra et al. 2012). 68\% of the GRBs of the BAT6 sample have radio observations (Table~\ref{tab1}). These were obtained, in most cases, with the VLA at 8.4 GHz through pointings between 1 and 5 days after the burst explosion. We verified that the sub--sample of the BAT6 with radio observations has the same properties of the BAT6 sample. Our radio selected sample represents an unbiased $\gamma$--ray selected sample and offers the unique opportunity to study the properties of GRB radio afterglows. We find that about half of the observed GRBs of the BAT6 sample were detected at radio frequencies. For the other half we collected the 3$\sigma$ upper limits (Table~\ref{tab1}). Radio detections have a median flux at $\sim$2 days of $\approx$100 $\mu$Jy with only 3-6\% brighter than 500 $\mu$Jy. 
The higher detection rate (50\%) of the radio afterglows of the BAT6 sample, compared to the $\sim$30\% found by CF12, is partly due to the fact that the BAT6 sample selects bright \sw\ GRBs (i.e. with a $\gamma$--ray peak flux $P\ge$ 2.6 ph cm$^{-2}$ s$^{-1}$). We note that this detection rate could be a lower limit due to the fact that the non detection of the GRB afterglow in late time observations often prevents the subsequent follow up thus missing bursts peaking at later times. This is particularly evident for $\gamma$--ray bright bursts as those of the BAT6 sample (see Fig.\ref{fg4}) 

We built a code, PSYCHE, by coupling a Population SYnthesis Code (Ghirlanda et al. 2013) with a Hydrodynamical Emission model for the afterglow (van Eerten et al. 2012). PSYCHE allows us to simulate the afterglow emission (through the Hydrodynamical Emission model) of the population of GRBs (simulated by the Population Synthesis Code). We restrict our interest to the population of GRBs pointing towards the Earth, i.e. the bursts that can trigger a  dedicated $\gamma$--ray detector.  The free assumptions of PSYCHE are the microphysical parameters at the shock front, responsible for the afterglow emission, described by the Hydrodynamical Emission part of the code. Indeed, the free parameters of the population synthesis code have already been set in Ghirlanda et al. (2013) to reproduce the high energy properties of the GRB populations detected by different satellites ({\it CGRO}, \sw\ and \fe). By assuming standard afterglow parameters (i.e. a constant density uniformly distributed between 0.1 and 30 cm$^{-3}$, $\epsilon_e=0.02$, $\epsilon_B=0.008$, $p=2.5$ - Panaitescu \& Kumar 2000; Ghisellini et al. 2009; Sironi et al. 2013) and a constant radiative efficiency $\eta=20$\% we can reproduce through PSYCHE the radio flux distribution of the BAT6 sample (Fig.~\ref{fg1}). 

Through PSYCHE we can derive the radio flux distributions at different epochs (Fig.~\ref{fg2} and Fig.~\ref{fg3}) of all the GRBs that can be detected from  Earth. In this population are present the simulated bright \sw\ GRBs, which are directly compared with the BAT6 sample in terms of radio flux distribution (Fig.~\ref{fg1}), but also the much larger population of GRBs weaker than the BAT6 flux limit, i.e. with peak flux $P<2.6$ ph cm$^{-2}$ s$^{-1}$. This allows us to make predictions on the detection rate at radio frequencies with future radio telescopes considering the possibility that a dedicated $\gamma$--ray detector like \sw/BAT (i.e. reaching a flux limit of $\sim$0.4 cm$^{-2}$ s$^{-1}$) or, hopefully, more sensitive than \sw/BAT will be available. 

Considering JVLA and MeerKAT2 (Table~\ref{tab2}), we predict that they can detect $\sim$66 GRB yr$^{-1}$ sr$^{-1}$  of the population of GRBs presently detected by \sw/BAT (i.e. with peak flux $\ge$0.4 cm$^{-2}$ s$^{-1}$). A similar rate is found for SKA2.  The major difference between present/forthcoming SKA precursors and the final SKA can be appreciated considering the entire population of GRBs. We forsee that SKA2 will be able to detect at radio frequencies all the bursts (with a detection rate of $\sim$480 GRBs yr$^{-1}$ sr$^{-1}$) that a $\gamma$--ray detector much sensitive than \sw/BAT can detect. 

The detection rates mentioned above consider a typical relatively short ($\sim$30 min) observation centered at 8.4 GHz  performed at  5 days after the trigger. Early time observations are fundamental to identify the GRB counterpart also at radio frequencies and to study the broad band GRB emission coupling radio data with higher frequencies observations (optical/NIR). Early time radio observations would provide a unique test for the standard afterglow model that predicts that both a forward shock developing into the ISM and a reverse shock traveling through the expanding jet should form. The reverse shock component should produce an early afterglow peak which has been observed in a few optical afterglows (e.g. Sari \& Piran 1999; Meszaros \& Rees 1999; Bloom et al. 2009; Yu, Wang \& Liang 2009). In the radio band, CF12 suggest that a marginal evidence of an early peak appears in the average light curve of the monitored bursts of their sample. 
This peak could be due to the reverse shock component which is expected to peak at $\sim$1 day (see e.g. Gomboc et al. 2009). 
Soderberg \& Ramirez--Ruiz (2003) interpret some radio flares as due to the reverse shock component, however  also variation in the ISM density profile can produce such flares. For such studies automated radio follow up facilities like the Arcminute Microkelvin Imager Large Array (AMI--LA, Staley et al. 2013) are fundamental and will likely provide a larger unbiased sample of radio observed GRBs.

The radio detection rates reported in Table~\ref{tab1} depend (in addition to the obvious scaling of the sensitivity with the exposure time, $t_{\rm exp}^{-1/2}$), on the observing frequency and time. At early times the radio afterglow emission is self--absorbed and the radio light curve should peak at relatively late times (1--10 days) when the optically thick to thin transition happens. We confirm (Fig.~\ref{fg3}) that at larger radio frequencies the radio afterglow peaks at earlier times and is brighter. This effect of the afterglow emission should be considered for the planning of the observation strategy. The high frequency observations at 22 GHz now possible with ATCA can detect the brightest afterglows peaking, between 0.5 and 10 days, with a peak flux of 0.1--1 mJy (bottom panel of Fig.~\ref{fg4}).  The low frequency domain (e.g. 1.4 GHz) will be covered by ASKAP and MeerKAT1. While ASKAP will detect a small fraction of GRBs at 1.4 GHz (top panel of Fig.~\ref{fg4}), MeerKAT1 can detect $\sim$80\% of the population of \sw\ burst at such frequencies if observations are performed at relatively late times, i.e. $\sim$10 days after the GRB explosion. 

Observations around the peak of the radio afterglow light curve offer the opportunity to measure circular and linear polarization, i.e. a powerful diagnostic of the physical conditions and of the emission process (thought to be synchrotron). At relatively high frequencies (i.e.  above the self--absorption frequency $\nu_a$) we expect to measure circular and linear polarization comparable to the level observed in the optical (Covino 2004; Lazzati 2006; Steele et al. 2010; Wiersama et al. 2012). Observations in the optically thick regime ($<\nu_a$) could show the effects of Faraday rotation or conversion (Toma et al. 2008) allowing an independent way to estimate the shock plasma parameters. 


Radio observations offer also a way to identify powerful explosions possibly arising from massive Pop--III stars. Indeed, in spite of their distance, these are predicted to have peak radio fluxes and peak times larger than the normal Pop--II GRB population. We have shown (Fig.~\ref{fg4} middle panel) that extremely energetic GRBs with $E_{\rm iso}=10^{55}$ erg, as those expected from Pop-III progenitors (e.g. Toma, Sakamoto \& Meszaros 2011) at high redshifts, should be extremely bright (10-30 mJy) at late times (violet symbol in Fig.~\ref{fg4} middle panel).  The detection of a bright radio source with a light curve peaking $\sim$100 days for a NIR dark long GRB could be the signature of a very energetic events exploding at $z\gsim 19$.

Across $t_{NR}$ the light curve temporal decay index should vary (either flattening or steepening - Frail 2004). By measuring this variation it is possible to put some constraints on the slope of the electron energy distribution at the shock front. Also alternative explanations for the flattening of the light curve at $t_{NR}$ are possible (e.g. time evolution of the microphysical shock parameters - Panaitescu 2000, Rossi \& Rees 2003 - or late time energy injection - Panaitescu et al. 1998) and the monitoring of the afterglow emission across $t_{NR}$ at different radio frequencies should disentangle among such different interpretations (e.g. late time energy injection should produce an achromatic light curve flattening). If the interstellar medium has a wind density profile (as it is  expected if the progenitor star underwent intense mass loss  before its collapse), $t_{NR}$ should happen later (and with  a different slope change). Therefore, the detection of the NR transition in the radio band would provide also a diagnostic of the ISM structure (either constant density or wind). 

Late time radio observations after the trans--relativistic transition will offer a unique opportunity to estimate the {\it true} kinetic energy of GRBs, $E_{k}$. This estimate has been possible, so far, only in a handful of cases (Frail et al. 2000; Berger et al. 2004, Frail et al. 2005) while lower/upper limits were recently derived for a relatively larger sample of events  (Shivvers \& Berger 2012). $E_{k}$ estimated from the late time radio observations represents the true GRB kinetic energy.

The relevance of estimating $E_{k}$ is twofold: (a) for the  GRBs with a measured jet opening angle \th, i.e. those with an estimate of the true prompt emission energy $E_{\gamma}\sim \theta_{jet}^{2} E_{\rm iso}$, it is possible to derive a unique estimate of the radiative efficiency $\eta\sim E_{\gamma}/E_{k}$ and, (b) if one assumes a typical efficiency $\eta$ through $E_{k}$ it is possible do derive an estimate of $E_{\gamma}$ independent from the collimation angle. This latter possibility would provide a large number of bursts with an estimate of $E_{\gamma}$ to be added to the $E_{\rm peak}-E_{\gamma}$ correlation  (Ghirlanda et al. 2004, Nava et al. 2006) and would provide a test for the use of GRBs to constrain the cosmological parameters (Ghirlanda et al. 2004a).

\section*{Acknowledgments}
We acknowledge D. Frail and P. Chandra for having provided the radio fluxes of two bursts. We thank P. Hancock for discussion. ASI I/004/11/0 and the 2011 PRIN-INAF grant are acknowledged for financial support. Development of the Boxfit code (van Eerten et al. 2012) was supported in part by NASA through grant NNX10AF62G issued through the Astrophysics Theory Program and by the NSF through grant AST-1009863.  We acknowledge the anonymous referee for comments.

\end{document}